# Transport properties of iron-based FeTe$_{0.5}$Se$_{0.5}$ superconducting wire

Toshinori Ozaki, Keita Deguchi, Yoshikazu Mizuguchi, Hiroaki Kumakura, and Yoshihiko Takano.

*Abstract*— FeTe$_{0.5}$Se$_{0.5}$ superconducting wires have been fabricated using ex-situ PIT method with an Fe sheath. Among the other FeAs-based superconductors, FeTe$_{0.5}$Se$_{0.5}$ has great advantage for applications due to its binary composition and less toxicity. Surprisingly, superconducting current was observed in the as-fabricated wire without any heat treatments. The zero-resistivity critical temperature ($T_c^{zero}$) and transport critical current density ($J_c$) at 2 K were 3.2 K and 2.8 A/cm$^2$, respectively. In addition, by annealing at 200°C for 2 h, $T_c^{zero}$ and $J_c$ were enhanced up to 9.1 K, and 64.1 A/cm$^2$, respectively. This suggests that the inter-grain connectivity was improved by heat treatment, and superconducting properties of FeTe$_{0.5}$Se$_{0.5}$ wire were enhanced.

*Index Terms*— FeTe$_{0.5}$Se$_{0.5}$; Superconducting wires; Critical temperature; Critical current density;

## I. INTRODUCTION

THE discovery of iron-based superconductors [1] inspired an immense research activity in this field. Over the past two years, several groups of iron-based superconductors have been discovered, such as BaFe$_2$As$_2$ (122 series) [2], LiFeAs (111 series) [3] and FeSe (11 series) [4]. Among them, The 11 series, such as FeSe, FeTe$_{1-x}$Se$_x$ and FeTe$_{1-x}$S$_x$, is an important ferrous superconducting system. The 11 series has the simplest structure without block layers along the c-axis, and less toxicity compared to the other FeAs-based superconductors [5]-[8]. Although the transition temperature ($T_c$) of FeSe is as low as 8.5 K, the $T_c$ can be highly enhanced to 15 K by partial substitution of Te or S [5]-[10], and up to 37 K under high pressure [11]-[13]. Moreover, 11 series have a high upper critical field ($H_{c2}$) [6], [14], [15]. Therefore, the 11 series have a great potential for applications.

We succeeded in the observation of a zero-resistivity current on the current-voltage measurement for the iron-based superconducting wire using the in-situ powder-in-tube (PIT) method with an Fe sheath [16]. Up to now, there are hardly any new reports on the fabrication of superconducting wires in which transport critical current density ($J_c$) values were obtained [16], [17]. Furthermore, obtained $J_c$ in these wires is much lower than intra-grain $J_c$ [18]. Lee *et al.* pointed out that grain boundaries in 122 series exhibit current-limiting behavior similar to that observed in high-$T_c$ cuprates [19]. However, heat treatment may enhance grain connectivity and result in $J_c$ improvement [20]. In this paper, we report on the fabrication of the FeTe$_{0.5}$Se$_{0.5}$ superconducting wire using the ex-situ PIT method with an Fe sheath, and the effect of heat treatment on superconductivity in FeTe$_{0.5}$Se$_{0.5}$ wire.

## II. EXPERIMENTAL

FeTe$_{0.5}$Se$_{0.5}$ wires were prepared using an ex-situ PIT method with an Fe sheath. First of all, polycrystalline samples of FeTe$_{0.5}$Se$_{0.5}$ were prepared using the solid state reaction method. High purity powders of Fe (99.9% up), Se (99.999%) and Te (99.999%) were mixed with nominal compositions and sealed into evacuated quartz tubes. Then the powders were heated at 650°C for 15 h. The obtained samples were ground and pressed into pellets. The pellets were reheated at 650°C for 15 h in evacuated quartz tubes. After furnace cooling, the product was reground and packed into a pure Fe tube with a length of 5 cm. The inner and outer diameters of the tube were 3.5 and 6 mm, respectively. The composite was initially fabricated in a rod shape with about 2.5 mm × 2.5 mm cross section by groove rolling and then drawing into a wire of 1.1 mm in diameter. Fig. 1(a) shows optical micrographs of the transverse cross section of the wire. The cross section shows

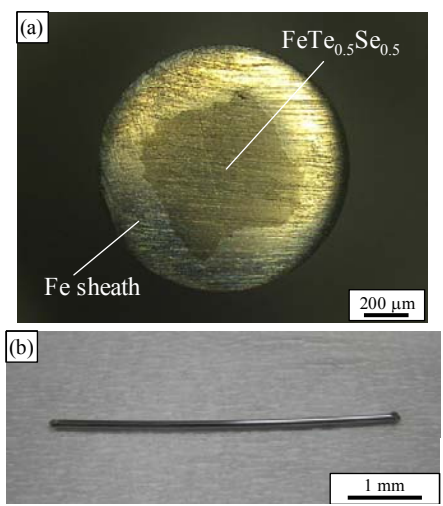

Fig. 1. FeTe$_{0.5}$Se$_{0.5}$ wire fabricated by the ex-situ PIT method. (a) Optical micrograph of the transverse cross section of wire after heat treatment at 200°C for 2 h. (b) Photograph of samples after cutting and before heating.

Manuscript received 3 August 2010. This work was supported by Japan Society for the Promotion of Science (22-10227). Part of the work was supported in part by Grant-in-Aid for Scientific Research (KAKENHI)

T. Ozaki, K. Deguchi, Y. Mizuguchi, H. Kumakura, and Y. Takano are with Superconducting Materials Center, National Institute for Materials Science, 1-2-1 Sengen, Tsukuba 305-0047, Japan, and JST-TRIP, Japan Science and Technology Agency-Transformative Research-Project on Iron-Pnictides, 1-2-1 Sengen, Tsukuba 305-0047, Japan, (phone: +81-29-859-2644; fax: +81-29-859-2601; e-mail: OZAKI.Toshinori@nims.go.jp, DEGUCHI.Keita @nims.go.jp,MUZUGUCHI.Yoshikazu@nims.go.jp,KUMAKURA.Hiroaki @nims.go.jp,TAKANO.Yoshihiko@nims.go.jp).

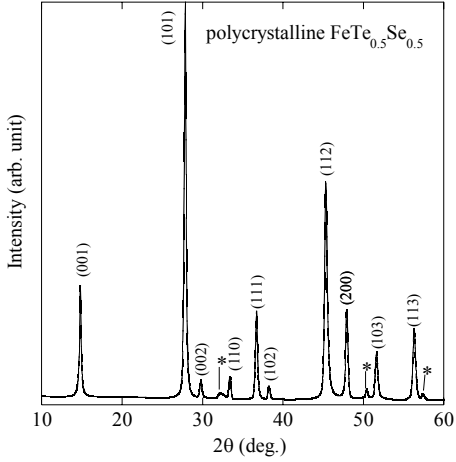

Fig. 2. Powder x-ray diffraction pattern for the as-grown $FeTe_{0.5}Se_{0.5}$ samples. The asterisks indicate the peaks of impurity phase.

uniform deformation of composite without any breakage. The round-shaped wire was cut into pieces of about 4 cm in length (Fig. 1(b)), and the wires were sealed into a quartz tube with an atmospheric-pressured argon gas. The sealed wires were rapidly heated at 150-500°C for 2 h.

Constituent phases were determined by powder X-ray diffraction (XRD) with Cu K$\alpha$ radiation and intensity data were collected in the $2\theta$ range of 5-60°. Temperature dependence of magnetic susceptibility was measured using a superconducting quantum interference device (SQUID) magnetometer at $H = 10$ Oe. Temperature dependence of resistivity and transport critical current density $J_c$ were measured by the four-probe method with a physical property measurement system (PPMS).

## III. RESULT AND DISCUSSION

### A. Polycrystalline $FeTe_{0.5}Se_{0.5}$

The x-ray diffraction pattern of $FeTe_{0.5}Se_{0.5}$ compound is shown in Fig. 2. All the peaks were well indexed using a space group of P4/*nmm* except for an unidentified small peak as an impurity phase labeled by asterisks. The compound crystallizes in a tetragonal structure and no secondary phase is observed. Lattice parameters are found to be $a$=3.7943(2) and $c$=5.9966(5) Å. The lattice parameters are in good agreement with previously published reports [5], [21].

Temperature dependence of zero-field-cooled (ZFC) and field cooled (FC) magnetization of polycrystalline $FeTe_{0.5}Se_{0.5}$ is shown in Fig. 3(a). The sample showed superconducting transition at 13.9 K. The superconducting volume fraction estimated from the ZFC magnetization at 2 K was larger than 100% due to the porous structure and a demagnetization effect of the sample. Fig. 3(b) shows the temperature dependence of the resistivity of $FeTe_{0.5}Se_{0.5}$ compound. Resistivity around $T_c$ is shown in the inset. While this compound has a somewhat semiconducting behavior above 150 K, it shows metallic behavior in the normal state below 150 K. Similar temperature dependence of resistivity is reported for samples with compositions close to $FeTe_{0.5}Se_{0.5}$ [7], [8], [21], [22]. This behavior has been attributed to a weak charge-carrier localization due to a large amount of excess Fe in $Fe_{1+y}Te_{1-x}Se_x$ system [23]. Additionally, the value of resistivity is higher than that reported in the previous reports [7], [8], [21], [22]. This could be resulted from the microscopic inhomogeneities that are still present in our sample and impurity phases. The superconducting transition was observed at $T_c^{onset}$ of 15.7 K and zero-resistivity was achieved at 12.8 K. These values are almost the same as previous reports on polycrystalline and single crystal $FeTe_{0.5}Se_{0.5}$ samples [19]-[23].

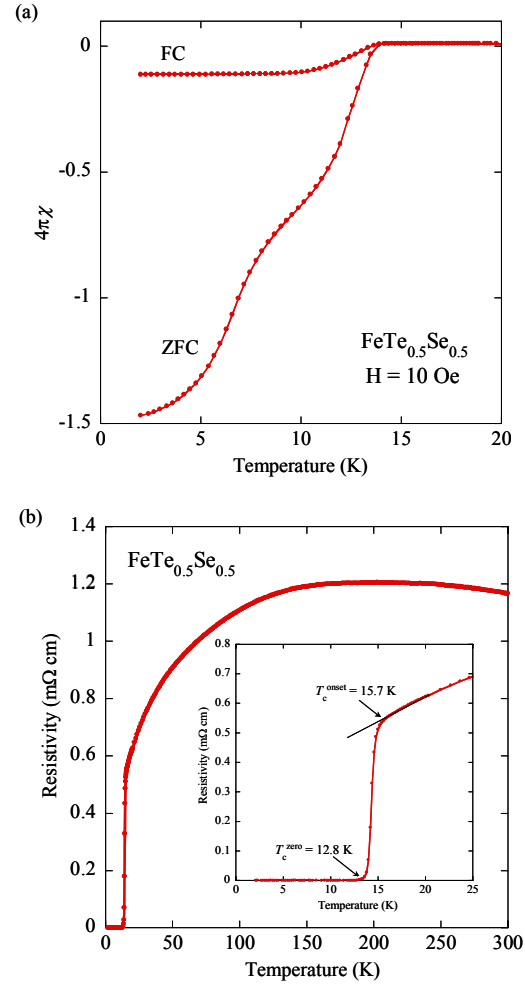

Fig. 3. (a) Temperature dependence of the zero-field-cooled (ZFC) and field-cooled (FC) magnetization at $H = 10$ Oe and (b) temperature dependence of resistivity of a sintered bulk sample with a nominal composition of $FeTe_{0.5}Se_{0.5}$.

### B. $FeTe_{0.5}Se_{0.5}$ superconducting wire

Fig. 4 shows the temperature dependence of resistivity at zero magnetic fields for $FeTe_{0.5}Se_{0.5}$ wires fabricated by ex-situ PIT method. The values of $T_c^{onset}$, $T_c^{zero}$ and transition width ($\Delta T_c$) are listed in table 1. Surprisingly, superconducting current was observed in the as-fabricated wire without any heat treatments. The zero-resistivity critical temperature ($T_c^{zero}$) and transport critical current density ($J_c$) at 2 K were 3.2 K and 2.8 A/cm$^2$, respectively. $J_c$ was obtained using the criteria of $E = 1$ $\mu$V/cm. In addition, with increasing temperature of heat treatment, the value of $T_c^{onset}$ and $T_c^{zero}$ increased. The highest $T_c^{onset}$, $T_c^{zero}$ and $\Delta T_c$ were 10.8 K, 9.1 K and 1.7 K for $FeTe_{0.5}Se_{0.5}$ wire annealed at 200°C for 2 h, respectively. The narrow $\Delta T_c$ is indicative of high homogeneity of the superconducting property. Note that the transport $J_c$ value showed 82.5 A/cm$^2$ at 2 K and 64.1 A/cm$^2$ at 4.2 K, a factor of 5 higher than the value obtained Fe (Se,Te) wire fabricated by in-situ PIT method [16]. It is considered that the heat treatment in $FeTe_{0.5}Se_{0.5}$ improved the inter-grain connectivity. However, with an increase in temperature above 250°C, the values of $T_c^{zero}$ were reduced. Furthermore, $FeTe_{0.5}Se_{0.5}$ wires annealed above 400°C did not show zero-resistivity. We recently confirmed that Fe sheath reasonably supplied Fe for synthesizing the superconducting phase of Fe (Se,Te) [16]. Moreover, it is reported that the superconducting transition in single crystal of $Fe_yTe_xSe_{1-x}$ was suppressed by excess Fe [23]. Given these results, Fe sheath might excessively supply Fe for the superconducting phase, and the increase of excess Fe concentration resulted in reduced superconductivity.

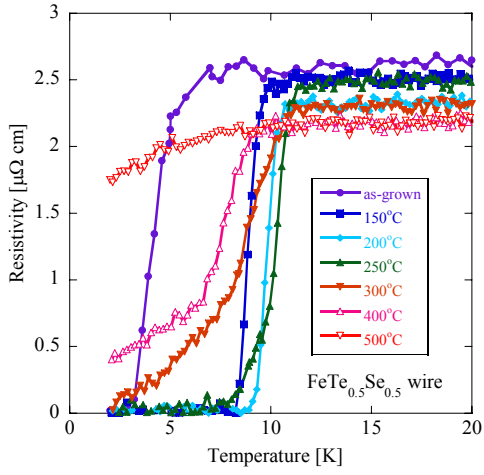

Fig. 4 Temperature dependence of resistivity for $FeTe_{0.5}Se_{0.5}$ superconducting wires at as-grown for different annealing temperatures $T$ = 150-500°C.

TABLE I $T_c^{ONSET}$, $T_c^{ZERO}$ AND $\Delta T_C$ FOR $FeTe_{0.5}Se_{0.5}$ SUPERCONDUCTING WIRES AT AS-GROWN AND DIFFERENT ANNEALING TEMPERATURES $T$ = 150-500°C.

| Annealing Temperature | $T_c^{onset}$ [K] | $T_c^{zero}$ [K] | $\Delta T_c$ [K] |
|---|---|---|---|
| As-grown | 6.8 | 3.2 | 3.6 |
| 150°C | 10.0 | 8.3 | 1.7 |
| 200°C | 10.8 | 9.1 | 1.7 |
| 250°C | 11.4 | 8.0 | 3.4 |
| 300°C | 11.0 | 2.4 | 8.6 |
| 400°C | 9.1 | — | — |
| 500°C | 7.2 | — | — |

Furthermore, it is notable that the $T_c^{onset}$ values of each $FeTe_{0.5}Se_{0.5}$ wire were suppressed more than 4 K compared to than that of $FeTe_{0.5}Se_{0.5}$ polycrystalline bulk. This result could be attributed to stress of $FeTe_{0.5}Se_{0.5}$ phase due to rolling process.

Fig. 5(a) shows the temperature dependence of the resistivity under various magnetic fields for the $FeTe_{0.5}Se_{0.5}$ wire annealed at 200°C for 2 h. With increasing magnetic fields, the $R(T)$

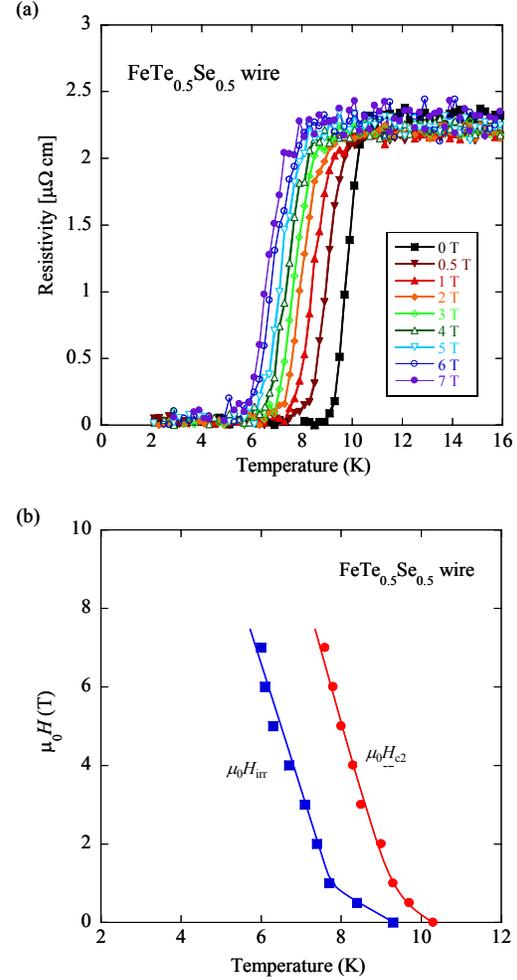

Fig. 5 (a) Temperature dependence of resistivity for $FeTe_{0.5}Se_{0.5}$ superconducting wires annealed at 200°C for 2 h under magnetic fields up to 7 T. (b) Upper critical field ($\mu_0H_{c2}$) line and Irreversibility field ($\mu_0H_{irr}$) line as a function of temperature for $FeTe_{0.5}Se_{0.5}$ superconducting wires annealed at 200°C for 2 h.

curves are shifted to lower temperature, but also they do not much broaden, indicating that a two-dimensional feature and the superconducting fluctuations in this wire are small. The upper critical field ($\mu_0H_{c2}$) and the irreversibility field ($\mu_0H_{irr}$) of the $FeTe_{0.5}Se_{0.5}$ wires annealed at 200°C for 2 h are plotted in Fig. 5(b) as a function of temperature. We have estimated $\mu_0H_{c2}$ and $\mu_0H_{irr}$, using 90% and 10% of normal state resistivity, respectively. The $\mu_0H_{c2}(0)$ is estimated to be ~40 T by linear extrapolation. Applying the Werthamer-Helfand-Hohenberg (WHH) theory [24], the $\mu_0H_{c2}(0)$ is calculated to be ~28 T. In order to approximate the superconducting parameters, we have used the Ginzburg-Landau formula for the coherence length ($\xi$), $\xi = (\Phi_0/2\pi\mu_0H_{c2})^{1/2}$, where $\Phi_0 = 2.07$ Oe cm$^2$, $\xi(0)$ was calculated as ~3.4 nm. Additionally, as shown in Fig. 5(b), $\mu_0H_{irr}$ values at 4.2 K and 0 K are estimated to be ~13 T and ~25 T, respectively, by linear extrapolation. In polycrystalline $FeTe_{0.5}Se_{0.5}$, the $\mu_0H_{c2}(0)$ ~120 T was reported [14]. We found that the heat treatment improved superconducting properties due to better connections between grains. Superconducting properties can be further enhanced by improvement of inter-grain connectivity. Therefore, $FeTe_{0.5}Se_{0.5}$ wires have a great potential for enhanced superconducting properties by optimization of the fabrication process.

## IV. CONCLUSION

We fabricated FeTe$_{0.5}$Se$_{0.5}$ superconducting wire using the ex-situ PIT method with an Fe sheath. In the as-grown FeTe$_{0.5}$Se$_{0.5}$ wire without heat treatment, zero-resistivity was observed at 3.2 K and transport $J_c$ showed 2.8 A/cm$^2$ at 2 K. In addition, by annealing at 200°C for 2 h, $T_c^{zero}$ and $J_c$ were enhanced up to 9.1 K and 82.5 A/cm$^2$ at 2 K, respectively. The heat treatment improved the superconducting properties such as $T_c$ and $J_c$ due to improvement of the inter-grain connectivity. For applications, it is desirable to improve the transport $J_c$ and $\mu_0 H_{irr}$. FeTe$_{0.5}$Se$_{0.5}$ has high intra-grain $J_c$ and high $\mu_0 H_{c2}$. Therefore, it is quite possible that significant improvement of transport $J_c$ and $\mu_0 H_{irr}$ can be achieved by the enhancement of inter-grain $J_c$.